\documentclass[11pt]{elsart}
\usepackage{amsmath,amssymb}
\usepackage{epsfig}
\usepackage{color}
\newcommand{\bv}{\boldsymbol{v}}

\newcommand{\bkhat}{\boldsymbol{\hat k}}
\newcommand{\bx}{\boldsymbol{x}}

\newcommand{\bz}{\boldsymbol{z}}

\newcommand{\Kn}{\mbox{Kn}}
\newcommand{\Ma}{\mbox{Ma}}

\begin{document}

\begin{frontmatter}

\title{A direct method for the Boltzmann equation based on a pseudo-spectral velocity space discretization} 

\author[1]{G. P. Ghiroldi},
\ead{gian.ghiroldi@mail.polimi.it}
\author[2]{L. Gibelli\corauthref{cor}}
\corauth[cor]{Corresponding author.}
\ead{livio.gibelli@polimi.it}

\address[1]{Politecnico di Milano, 
            Dipartimento di Matematica,
            Piazza Leonardo da Vinci 32, 
            20133 Milano, Italy}
\address[2]{Politecnico di Milano, 
            Dipartimento di Scienze e Tecnologie Aerospaziali,
            Via La Masa 34, 
            20156 Milano, Italy}

\begin{abstract}
A deterministic method is proposed for solving the Boltzmann equation.
The method employs a Galerkin discretization of the velocity space and adopts,
as trial and test functions, the collocation basis functions based on weights and roots 
of a Gauss-Hermite quadrature. This is defined by means of half- and/or full-range Hermite polynomials
depending whether or not the distribution function presents a discontinuity in the velocity space. 
The resulting semi-discrete Boltzmann equation is in the form of a system of hyperbolic partial differential
equations whose solution can be obtained by standard numerical approaches.  
The spectral rate of convergence of the results in the velocity space is shown by solving 
the spatially uniform homogeneous relaxation to equilibrium of Maxwell molecules. 
As an application, the two-dimensional cavity flow of a gas composed by hard-sphere molecules is studied 
for different Knudsen and Mach numbers.
Although computationally demanding, the proposed method turns out to be an effective tool for studying  
low-speed slightly rarefied gas flows. 
\end{abstract}


\end{frontmatter}

% ==============================================================================

\section{Introduction}

The conventional continuum approach to gas dynamics, namely the Navier-Stokes equations with no-slip boundary conditions, 
is justified when the {a\-ve\-ra\-ge} distance traveled by molecules between two successive collisions, $\lambda$, is much 
smaller than a characteristic length, $L$, associated to the flow geometry.   
This condition breaks down in several physical situations ranging from the re-entry of spacecraft in upper planetary atmospheres, 
characterized by large $\lambda$, to fluid-structure interaction in small-scale micro-electro-mechanical systems, characterized by small $L$.
In such situations, a microscopic description of the gas based on the Boltzmann equation is required~\cite{c88}.
In the absence of external forces, the Boltzmann equation for a gas composed by a single monatomic species whose atoms have mass $m$,  
takes the form
\begin{equation}
\label{eq:BE}
\frac{\partial f}{\partial t}+\bv\cdot\nabla_{\bx}f=\mathcal{C}(f,f) 
\end{equation}
where
\begin{equation}
\label{eq:Collint}
\mathcal{C}(f,f)=\int_{\mathbb{R}^3} \int_{\mathcal{S}^2} 
                 \left[f^* f_{1}^*-f f_{1} \right]\sigma(\|\bv_r\|,\bkhat\cdot \bv_r)\|\bv_r\| d\bv_1 d^2\bkhat
\end{equation}
In Eqs. (\ref{eq:BE},\ref{eq:Collint}), $f(\bx,\bv,t)$ denotes the distribution function of atomic velocities $\bv$ at spatial location $\bx$ and time $t$, 
whereas $\mathcal{C}(f,f)$  gives the collisional rate of change of $f$ at the phase space point $(\bx,\bv)$ at time $t$
and we have used the shorthand $f^{*}=f(\bx,\bv^{*},t)$, $f_{1}^{*}=f(\bx,\bv_{1}^{*},t)$ and $f_{1}=f(\bx,\bv_{1},t)$.
As is clear from Eq. (\ref{eq:Collint}), $\mathcal{C}(f,f)$ is a non-linear functional 
of $f$, whose precise structure depends on the assumed atomic interaction forces through the differential cross section 
$\sigma(\|\bv_r\|,\bkhat\cdot \bv_r)$.  
The dynamics of binary encounters determines the differential cross section as a function of the modulus $\|\bv_r\|$ of the relative velocity $\bv_r=\bv_1-\bv$ of two colliding atoms and of the orientation of the unit impact vector 
$\bkhat$ with respect to $\bv_r$. The collisional dynamics also determines the pre-collisional velocities $\bv^*$ and $\bv_{1}^*$
which are changed into $\bv$ and $\bv_{1}$ by a binary collision 

\begin{eqnarray}
\bv^*&=& \bv+(\bv_r\cdot \bkhat)\bkhat \\
\bv_{1}^*&=& \bv_{1}-(\bv_r\cdot \bkhat)\bkhat
\end{eqnarray}

\noindent
The prevalent approach for numerically solving Eq.~\eqref{eq:BE} is a stochastic-based method called Direct Simulation Monte Carlo (DMSC)~\cite{b94}.
The distribution function is represented by a number of mathematical particles which move in the computational domain and collide according to 
stochastic rules derived from Eqs.~\eqref{eq:BE}-\eqref{eq:Collint}.
Macroscopic flow properties are obtained by time averaging particle properties. If the averaging 
time is long enough, then accurate flow simulations can be obtained by a relatively small number of particles. 
Variants of DSMC have been proposed over the years to improve solution accuracy in the presence of high density gradients~\cite{rw96} and
for low Mach number flows~\cite{hh07}. Although particle-based methods are by far the most effective tools in describing non-equilibrium gas flows, 
they are not well suited to simulate unsteady gas flows.
Indeed, in this case the possibility of time averaging is lost or reduced. Acceptable accuracy can only be achieved by increasing the 
number of simulation particles or superposing several flow snapshots obtained from statistically independent simulations of the same flow but, in both cases, the 
computing effort is considerably increased. The simulation of steady gas flows in the near continuum limit represents an additional
challenge since the time scale on which the particle-based methods are forced to operate is much shorter than the characteristic macroscopic time 
and therefore explicit integration to steady state is computationally demanding.
Approaches based on a direct discretization of the Boltzmann equation in the phase space 
are believed to be a feasible alternative in these cases since they provide solutions with high accuracy even in unsteady conditions
and offer the possibility of a direct steady-state 
formulation~\cite{ar01,dpr04}. As such, they have been applied to study several problems of both theoretical interest and practical importance, including   
the viscous gas damping in microfluidic devices~\cite{lgffc07,cffggl08}, the onset of instability in a rarefied gas environment~\cite{ar11} and 
the investigation of ghost effects~\cite{sd04}. 
Deterministic methods of solution present some further assets compared to particle-based methods. 
Firstly, they are more suited to be adopted within a domain decomposition approach since the need to exchange information between kinetic and macroscopic equations 
requires smooth numerical solutions~\cite{ddm10,ap11}. 
Secondly, unlike particle-based methods
their implementation on massively parallel computers with SIMD architecture, such as multi-core and GPU processors, can easily realize the full potential of these 
supercomputers~\cite{fgg11a,fgg11b,hg12}.
These aspects also prompted the development of deterministic methods of solution.

\noindent
Considerable progress has been accomplished in developing deterministic {me\-thod} for kinetic model equations~\cite{m00,a11}. 
By contrast, an accurate and efficient direct solution of the Boltzmann equation itself remains a challenging problem. 
A common strategy {a\-dop\-ted} for solving Eq.~\eqref{eq:BE}
consists in decoupling the transport and the collision terms by time-splitting the evolution operator into a transport step and a collisional step.
The transport step requires to solve a system of hyperbolic conservation laws coupled at the boundaries.
Their discretization can be performed in a variety of ways, including finite-difference, finite-volume, finite-element or spectral methods~\cite{l02}. 
The collision step consists of solving a spatially homogeneous relaxation equation. This is the more computationally demanding part since it involves the 
computation of the bilinear five-fold integral defining the collision operator, Eq.~\eqref{eq:Collint}.
The numerical approaches to evaluate the collision step may be grouped into two broad categories. 
To the first category belong methods referred to as discrete velocity models (DVM). 
They make use of a Cartesian grid in velocity space and construct a discrete collision mechanics on the
points of the grid that preserve the main physical properties of the collision integral, namely
equilibrium states, collision invariants and entropy inequality~\cite{ph02}.
DVM methods have high computational cost and low order of accuracy although
fast algorithms have been recently developed for a restricted set of collision kernels~\cite{mpr12}.
To the second category belong methods which adopt a Galerkin discretization of the velocity space.
They are based on expanding the velocity dependence of the distribution function in a set of trial functions with expansion coefficients that depend on position 
and time. The Galerkin ansatz is substituted in the space homogeneous relaxation equation which is subsequently multiplied by test functions and integrated in the 
velocity space. According to the Galerkin approach, test and trial functions are assumed to be the same.
The above procedure yields a system of hyperbolic partial differential equations for the expansion coefficients.
Galerkin methods can be further distinguished depending on the basis functions which they employ. 
In Fourier-Galerkin approach, the distribution function is expanded in trigonometric polynomials and the fast Fourier Transform is used to
accelerate the computation of the collision integral in the velocity space. Several different methods have been developed starting from different representation of the
collision integral~\cite{pr00,mp04,fmp06,gt09}. These methods are generally very efficient and spectrally accurate for smooth solutions.  
Their major shortcoming is the loss of some of the properties of the solution such as positivity and conservation of momentum and energy. 
Preservation of collision invariants can be enforced but the use of corrections procedure may limit the accuracy of the solutions. 
Discontinuous Galerkin methods adopt discontinuous piecewise polynomials as test and trial functions~\cite{bh08,m11,aj12,aj12_b}.
Although computationally demanding, these methods have the remarkable feature to provide spectral accuracy in the velocity space even for 
discontinuous solutions which typically occur in the presence of solid surfaces. 
An hybrid approach is adopted in Refs.~\cite{o92,soa89} where the distribution function is expanded in Laguerre polynomials with respect to the velocity components 
parallel to solid surfaces whereas quadratic finite element functions have been used for the normal velocity component.  
This approach permits to explicitly account for the discontinuity of the distribution function and its application to one-dimensional problems
has been shown to provide accurate numerical solutions in both linearized and non-linear regime.

\noindent  
The main objective of the present work is to propose a new approach to the deterministic solution of the Boltzmann equation. 
The method uses a Galerkin discretization of the velocity space and adopts, as trial and test functions, 
the collocation basis functions based on weights and roots of a Gauss-Hermite quadrature.
The Boltzmann equation is thus simplified to a system of hyperbolic conservation laws with non-linear source terms in the physical space. 
A fully discrete numerical scheme can then be derived by using standard approaches~\cite{l02}. 
The method proposed in Ref.~\cite{gwc06} shows strong similarities with the present work but it differs in two significant respects.
In Ref.~\cite{gwc06}, the method is developed for a system of linear Boltzmann equations and
the collocation basis functions have been defined only on the basis of full-range Hermite polynomials.
By contrast, we are here concerned with the non-linear Boltzmann equation and the roots and weights of the half-range Gauss-Hermite
quadrature formula also enter in the definition of the collocation basis functions.
Half-range Hermite polynomials have been widely used to study flow and heat transfer problems in rarefied gas dynamics~\cite{gjz57,s03,fgf09,g12} and 
the Milne problem in radiative transfer~\cite{ls83}. In comparison with full-range Hermite polynomials,
they allow to deal exactly with boundary conditions and to achieve a faster convergence rate in the presence of discontinuous distribution functions. \\
The rest of the paper is organized as follows.
Section~\ref{sec:II} illustrates the method of solution in detail.
Section~\ref{sec:III} covers two aspects. Firstly, the spectral accuracy of the proposed velocity space discretization is illustrated by a comparison with the 
exact solution of the spatially homogeneous relaxation for Maxwellian molecules. 
Secondly, the possibilities of the proposed method are demonstrated by solving the two-dimensional driven cavity flow of a gas composed by hard-sphere molecules for
different Knudsen and Mach numbers.
Section~\ref{sec:IV} summarizes the main results and the future research directions.

\section{Mathematical formulation}
\label{sec:II}
The numerical scheme to deterministically solve the Boltzmann equation is derived through a two-step procedure. 
As a first step, Eq.~\eqref{eq:BE} is rewritten in terms of the deviational part of the distribution function, $h(\bx,\bv,t)$, defined as
\begin{equation}
\label{eq:definition}
 f(\bx,\bv,t)=M(\bv) \left[1+h(\bx,\bv,t)\right]
\end{equation}
where  $M(\bv)$ is the Maxwellian at equilibrium with uniform and constant density
$n_{0}$ and temperature $T_{0}$, i.e.,
 \begin{equation}
\label{eq:MaxwellDist}
M = \frac{n_{0}}{\left(2\pi R T_{0} \right)^{3/2}}
\exp\left(-\frac{\bv^2}{2RT_{0}}\right)
\end{equation}
As discussed in Refs.~\cite{fgg11b,bh08,aj12}, the use of $h$ permits to greatly improve the accuracy of the method of solution
when the gas approaches equilibrium since it reduces the truncation errors which arise in evaluating
Eq.~\eqref{eq:Collint} for distribution functions close to Maxwellian. In this respect, an equilibrium distribution function
which depends on space and time, $M(\bx,\bv,t)$, may be also adopted in Eq.~\eqref{eq:definition} to further increase accuracy of the 
computations. The applicability of the method, however, is not affected by this choice.  
Substituting Eq.~\eqref{eq:definition} into Eq.~\eqref{eq:BE} yields
\begin{multline}
\label{eq:devBoltz}
\frac{\partial h}{\partial t} + \bv \cdot \nabla_{\bx} h =  \int_{\mathbb{R}^3} \int_{\mathcal{S}^2} 
M_1 \left[h^*+h^*_1-h-h_1+\left(h^* h^*_1-hh_1\right)\right] \\
\sigma(\|\bv_r\|,\bkhat\cdot \bv_r)\|\bv_r\| d\bv_1 d^2 \bkhat
\end{multline}
It is worth noticing that if the perturbation is sufficiently small, i.e., $h \rightarrow 0$,
the quadratic terms in $h$ become negligible and Eq.~\eqref{eq:devBoltz}
simplifies to the linearized Boltzmann equation. 
The present formulation in terms of the deviational part of the distribution function, however,
is not restricted to a vanishing perturbation and it is valid in the non-linear case as well. \\
As a second step, we seek an approximate solution of Eq.~\eqref{eq:devBoltz} by using a Galerkin approach.
A suitable finite dimensional linear space of orthonormal trial functions 
$\phi_i(\bv)$ is chosen and the deviational part of the distribution function is expanded in a series of the form

\begin{equation}
\label{eq:h_N}
 h (\bx,\bv,t) \simeq h_{N} (\bx,\bv,t) = \sum_{j=0}^{N-1} h_j(\bx,t) \phi_j(\bv)
\end{equation}
Inserting Eq.~\eqref{eq:h_N} into Eq.~\eqref{eq:devBoltz} implies a residual depending on the expansion coefficients, $h_j (\bx,t)$.
These are determined by requiring that the residual is orthonormal with respect to some test functions. 
According to the Galerkin method, trial and test functions are assumed to be equal. 
Therefore, we substitute the spectral Galerkin ansatz, Eq.~\eqref{eq:h_N}, into Eq.~\eqref{eq:BE}, multiply both side by $\phi_i(\bv)$ and integrate
with respect to the velocity variable $\bv$. We thus obtain a closed first-order hyperbolic system of conservation laws
\begin{equation}
\label{eq:h_galerkin}
\frac{\partial h_i}{\partial t} + \sum_{\alpha=1}^{3} \sum_{j=0}^{N-1} C_{ij\alpha} \frac{\partial h_j}{\partial x_\alpha} = \sum_{j=0}^{N-1} L_{ij} h_j + 
                                                               \sum_{j,k=0}^{N-1} Q_{ijk} h_j h_k 
\end{equation}
In Eq.~\eqref{eq:h_galerkin}, the coefficient matrices $C_{ij\alpha}$, $L_{ij}$ and $Q_{ijk}$ are given, respectively, by
\begin{eqnarray}
\label{eq:int_a} 
C_{ij\alpha} & = & \int_{\mathbb{R}^3} M(\bv) \phi_i(\bv) \phi_j(\bv) v_\alpha d\bv \\
\label{eq:int_b}
L_{ij} & = & \int_{\mathbb{R}^3} \int_{\mathbb{R}^3} M(\bv) M(\bv_1) \,
\mathcal{K} \left( \phi_i,\bv,\bv_1 \right) \, \| \bv_r\| \, \phi_j (\bv) \, d\bv d\bv_1 \\
\label{eq:int_c}
Q_{ijk} & = & \frac{1}{2}\int_{\mathbb{R}^3} \int_{\mathbb{R}^3} M(\bv) M(\bv_1) \,
\mathcal{K} \left( \phi_i, \bv,\bv_1 \right) \, \| \bv_r\| \, \phi_j(\bv) \phi_k(\bv_1) \, d\bv d\bv_1 
\end{eqnarray}
The time-independent kernel of the integral operators in Eqs.~\eqref{eq:int_b} and~\eqref{eq:int_c},
$\mathcal{K}\left(\phi_i,\bv,\bv_1 \right)$, read,
\begin{equation}
\label{eq:kernel1}
\mathcal{K}\left(\phi_i,\bv,\bv_1 \right) =   \int_{\mathcal{S}^2}
                    \left[\phi_i(\bv^*) + \phi_i(\bv^*_1)-\phi_i(\bv)-\phi_i(\bv_1)\right] 
                    \sigma(\|\bv_r\|,\bkhat\cdot \bv_r)\, d^2 \bkhat 
\end{equation}
Boundary conditions can be treated similarly. For the sake of simplicity, let us consider a fixed plane wall at $x_2=0$ with temperature $T_0$. Let us further
suppose that the gas fills the half space $x_2>0$ and molecules which strike the wall are re-emitted according to the Maxwell's scattering kernel with 
complete accommodation~\cite{c88}.
In terms of the deviational distribution function, the boundary condition reads
\begin{equation}
\label{eq:bcMax}
h(0,x_2,x_3,\bv) = -\frac{1}{n_0} \sqrt{\frac{2\pi}{RT_0}}  \int_{\tilde{v}_2<0} M \tilde{v}_2 
                                                                   h(0,x_2,x_3,\tilde{\bv}) d\tilde{\bv}, \; \; \; v_2>0
\end{equation}
Substituting Eq.~\eqref{eq:h_N} into Eq.~\eqref{eq:bcMax}, multiplying both side by $\phi_i(\bv)$ and integrating with respect to the velocity variable $\bv$, yields
\begin{equation}
 h_i = \sum_{j=0}^{N-1}B_{ij} h_j
\end{equation}
where
\begin{equation}
\label{eq:int_d}
 B_{ij}  = -\frac{1}{n_0}\sqrt{\frac{2\pi}{RT_0}} \int_{\mathbb{R}^3}M(\tilde{\bv})\, \phi_i(\tilde{\bv})\, d\tilde{\bv} \; 
            \int_{\tilde{v}_2<0} M(\tilde{\bv})\, \phi_j(\tilde{\bv})\, \tilde{v}_2\, d\tilde{\bv} 
\end{equation}

\noindent
Several methods of solution for the Boltzmann equation employ a Galerkin discretization of the velocity space but
very different numerical schemes are derived due to the different trial and test functions~\cite{pr00,mp04,fmp06,gt09,bh08,m11,aj12,aj12_b,o92,soa89}.
In the present work, we adopt the collocation basis functions based on weights, $w_i$, and roots, $\bz_i$, of the Gaussian quadrature
formula defined by means of half- and/or full-range Hermite polynomials, i.e., $\phi_i(\bv)=\mathcal{B}_i(\bv)$ with $\mathcal{B}_i$ as given 
by Eq.~\eqref{eq:appendix_basis} (see Appendix~A for further details). 
A first consequence of this choice is that, by virtue of Eq.~\eqref{eq:appendix_collocation_property}, expansion coefficients are proportional to the values 
of the deviational part of the distribution function at the quadrature roots $\bz_i$, that is $h_i(\bx,t) = \sqrt{w_i} h_{N}(\bx,\bz_i,t)$. 
The proposed method of solution is thus closely related to discrete velocity methods.
Furthermore, computing the integrals over the velocity space in Eqs.~\eqref{eq:int_a}-\eqref{eq:int_c} as finite summations according 
to the Gauss-Hermite formula for numerical quadrature, Eq.~\eqref{eq:appendix_quadrature}, yields

\begin{eqnarray}
\label{eq:matrix_A}
& & C_{ij\alpha} = \delta_{i,j} z^{(\alpha)}_{i_\alpha} \\
\label{eq:matrix_B} 
& & L_{ij} =  \sqrt{w_j} \sum_{m=0}^{N-1} w_m \, \|\bz_j - \bz_m\| \, \mathcal{K}\left(\phi_i, \bz_j , \bz_m \right) \\
\label{eq:matrix_C}
& & Q_{ijk} =  \frac{1}{2}\sqrt{w_j w_k} \, \|\bz_j - \bz_k\| \, \mathcal{K}\left(\phi_i, \bz_j , \bz_k \right) 
\end{eqnarray}
Likewise, integrals in Eqs.~\eqref{eq:int_d} can be rewritten as

\begin{equation}
\label{eq:matrix_D}
 B_{ij}  = -\frac{1}{n_0}\,\sqrt{\frac{2\pi}{RT_0}} \sqrt{w_i} \; 
\sum_{z^{(2)}_{k}<0} \sqrt{w_k}\, \delta_{j,k} \, z^{(2)}_{k}
\end{equation}
Matrices given by Eqs.~\eqref{eq:matrix_B},\eqref{eq:matrix_C} and ~\eqref{eq:matrix_D}, can be computed with good accuracy once and for all and 
used in many individual simulations as long as their velocity space discretization is the same.
It is worth noticing that Gauss-Hermite quadrature formula can integrate exactly polynomials of degree at most $2N_\alpha-1$ in each velocity component $v_\alpha$.
Since integrals in Eqs.~\eqref{eq:int_b} and~\eqref{eq:int_c} may involve polynomials of higher order depending on the differential cross section,
their numerical evaluation is approximated.
The proposed method for the numerical solution of the Boltzmann equation has therefore a pseudo-spectral nature. \\
The collocation basis functions introduced above hold several advantages compared with alternative choices of functions $\phi_i$. 
Firstly, as shown by Eq.~\eqref{eq:matrix_A}, they lead to a diagonal system of conservation laws. This simplifies the problem of solving the 
advection step since boundary conditions are explicitly identified. 
Secondly, using half-range Hermite polynomials permits to exactly compute the half-space integrals appearing in the boundary conditions,
Eq.~\eqref{eq:int_d}. This permits to greatly improve accuracy of the solution for boundary value problems~\cite{gjz57,fgf09}.
It is worth pointing out that the use of half-range Hermite polynomials is not restricted to a cartesian geometry since, by making use of a  
body fitted coordinate system, arbitrary geometries can be dealt with as well~\cite{lz04}.
Finally, as it will be shown in Subsec.~\ref{sec:IIIsub1}, collision invariants are preserved during the relaxation step 
within machine precision and no correction method to enforce conservation of momentum and energy is thus required.

\section{Results and discussion}
\label{sec:III}

\begin{figure}[t]
\centering
\includegraphics[width=0.8\textwidth]{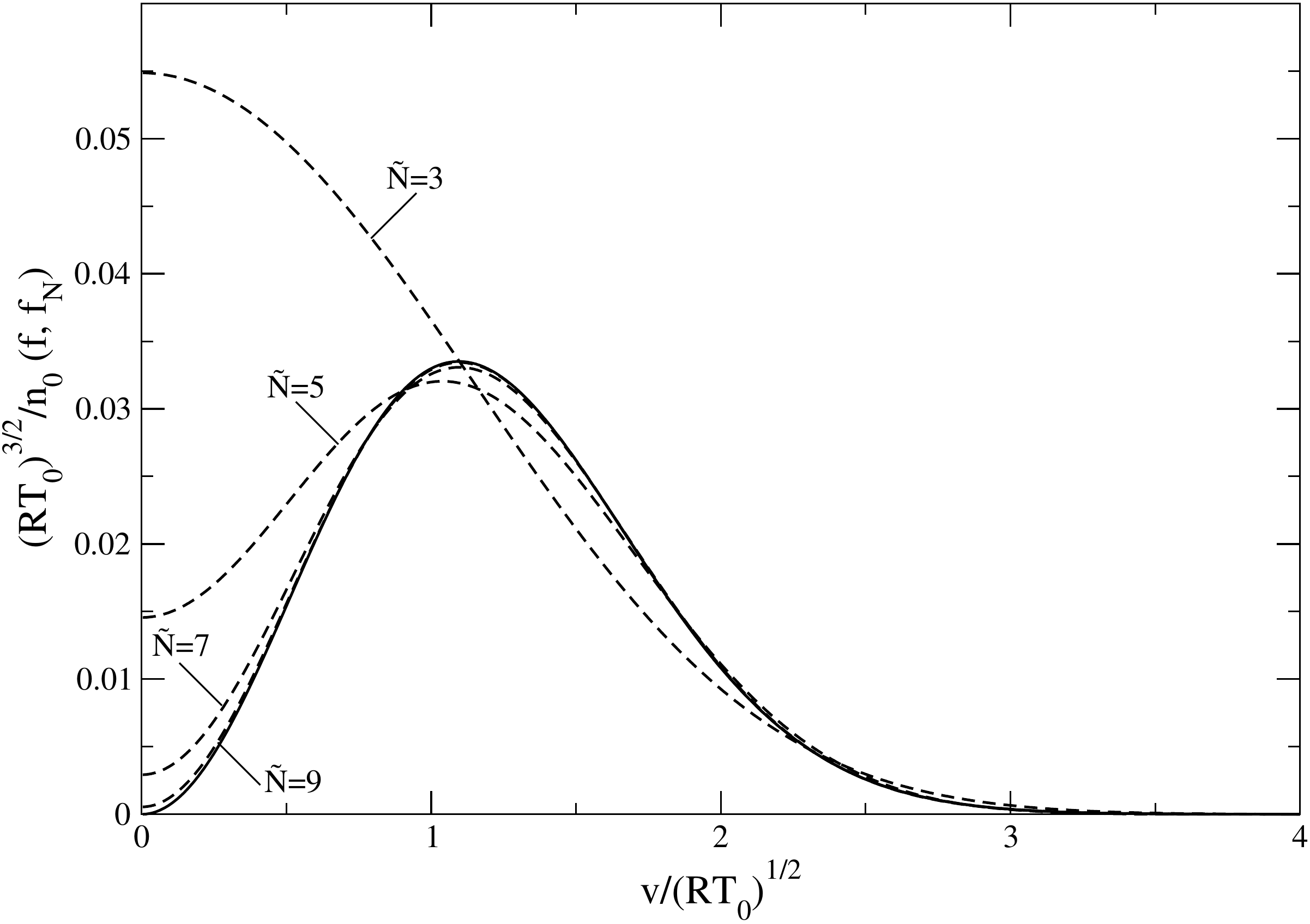}
\caption{Comparison between the BKW distribution function, Eq.~\eqref{eq:BKW_exact}, evaluated at $t=\bar{t}$ (solid line) and
         numerical distribution function, $f_N$, for different number of discrete velocities along each direction of the velocity space, $\tilde{N}$ (dashed lines).}
\label{fig:f_initial_condition}
\end{figure}

The key feature of the method described in Section~\ref{sec:II} is the introduction of a grid in the velocity space
whose points are the roots of the Gaussian quadrature formula based on half- and/or full-range Hermite polynomials. 
This discretization of the velocity space has major impact on the evaluation of the collision integral.
In Subsec.~\ref{sec:IIIsub1} the proposed method is thus assessed by computing spatially homogeneous relaxation for Maxwell molecules.
Most of the spectral approaches for solving the Boltzmann equation have been developed for 
a restricted set of collision kernels which do not include Maxwell molecules in the three-dimensional velocity space. By contrast, the
method proposed here can be carried over to arbitrary particle interactions and therefore we can consider 
the fully three-dimensional relaxation process. 
The comparison with an exact solution will show that mass, momentum and energy are conserved and numerical results are spectrally accurate 
in the velocity space. \\
In Subsec.~\ref{sec:IIIsub2} the possibilities of the proposed method are demonstrated by solving the driven cavity flow of a gas composed by 
hard-sphere molecules. This is a classical benchmark problem which it is often used for the validation of numerical codes.
In spite of its simple geometry, in fact, it contains most of the features of more complicated problems described by kinetic equations. 
A comprehensive study of this problem has been carried out in Refs.~\cite{nv05} for the Bhatnagar-Gross-Krook-Welander kinetic model 
equation in the linearized regime. By contrast, only few results are available for the hard-sphere Boltzmann equation~\cite{fgg11b,mesbr07}.
Subsec.~\ref{sec:IIIsub2} is thus of interest in that it provides accurate reference solutions albeit for a restricted range of Knudsen and Mach numbers.

\subsection{Spatially homogeneous relaxation}
\label{sec:IIIsub1}

\begin{figure}[t] 
\centering
\includegraphics[width=0.8\textwidth]{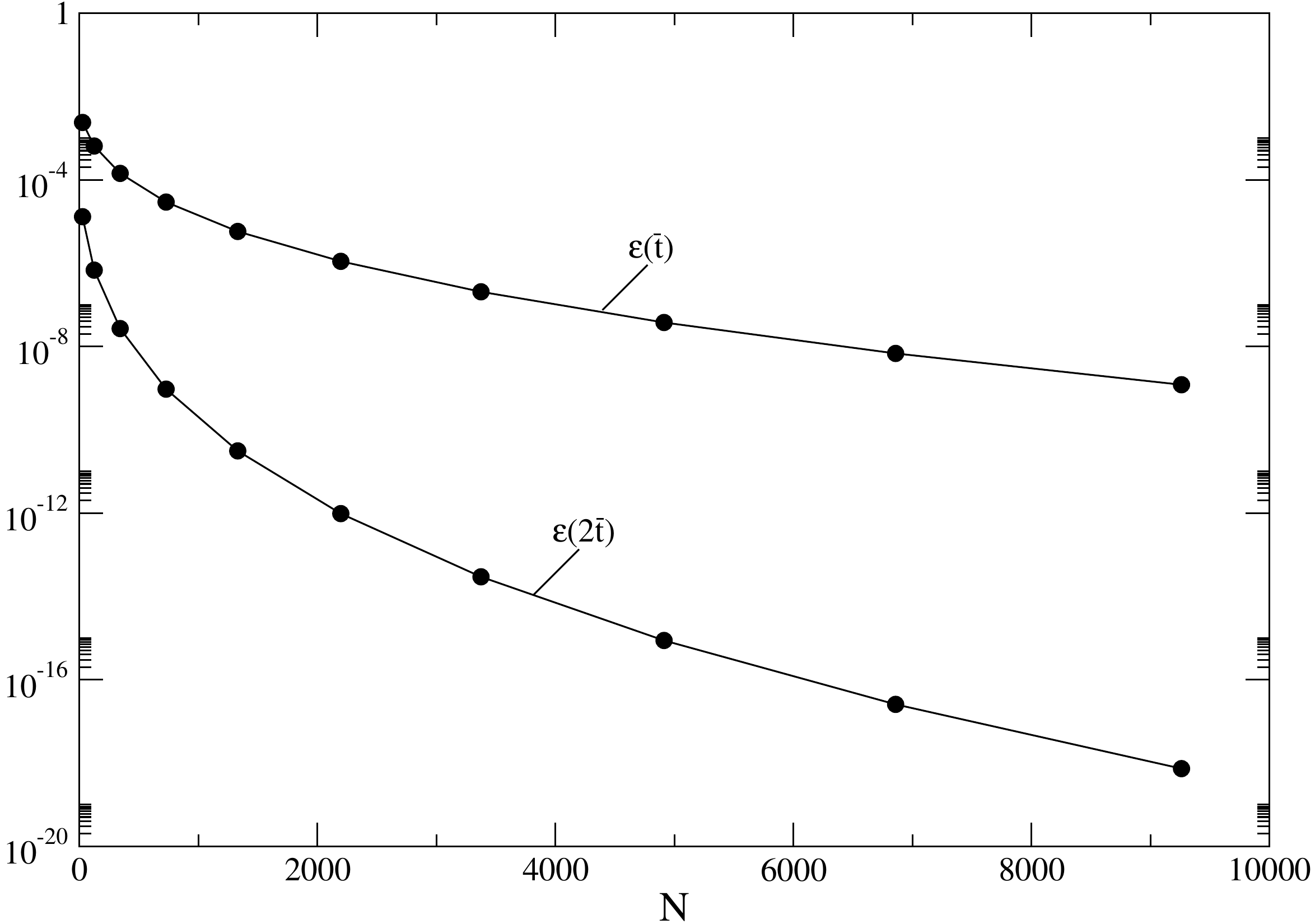}
\caption{Error between BKW and numerical distribution function, Eq.~\eqref{eq:error}, versus the total number of discrete velocities, $N$,
         for two different initial instants of time, $t_0=\bar{t}$ and $t_0=2\bar{t}$.}
\label{fig:error_initial_condition}
\end{figure}
\begin{figure}[t]
\centering
\includegraphics[width=0.8\textwidth]{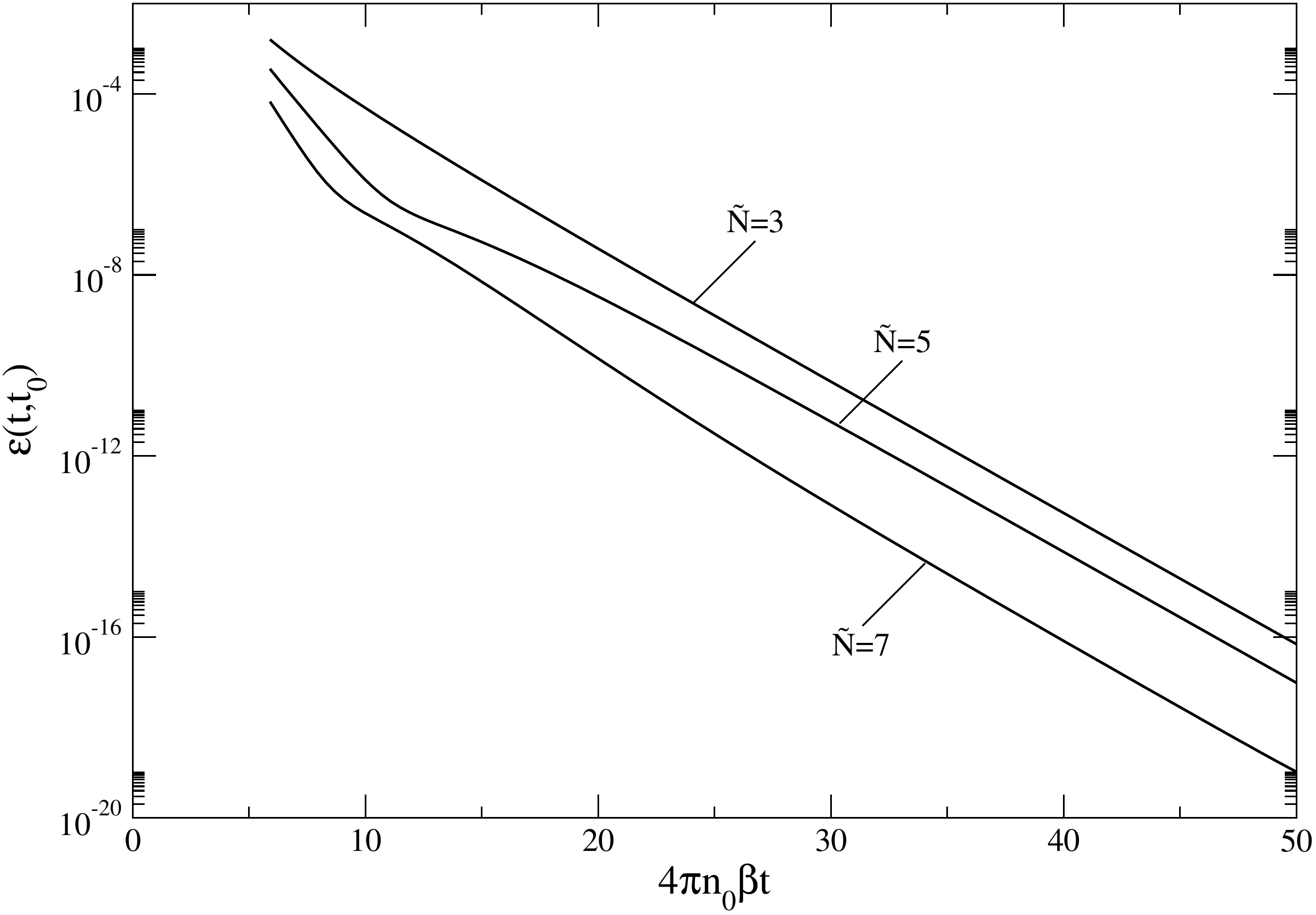}
\caption{Error between BKW and numerical distribution function, Eq.~\eqref{eq:error}, versus dimensionless time for different number of discrete velocities along each
         direction of the velocity space, $\tilde{N}$.}
\label{fig:error_vs_t}
\end {figure}

The method described in Sec.~\ref{sec:II} is here assessed by comparing the numerical solution of
the three-dimensional spatially homogeneous relaxation of Maxwell molecules 
with the exact Bobylev-Krook-Wu (BKW) distribution~\cite{b75,b88,kw77}.
The BKW distribution is radially symmetric and reads

\begin{equation}
 \label{eq:BKW_exact}
  f(v,t) = \frac{n_0}{\left[2\pi\alpha(t)RT_0\right]^{3/2}} \exp{\left(-\frac{v^2}{2RT_0\alpha(t)}\right)} 
           \left[ \frac{5\alpha(t)-3}{2\alpha(t)}+\frac{1-\alpha(t)}{2\alpha(t)^2}\frac{v^2}{RT_0}\right]
\end{equation}
where $n_0$ and $T_0$ are the equilibrium density and temperature, respectively, and 
\begin{equation}
 \alpha(t)=1-\exp{\left(-\frac{8}{3}\pi n_0 \beta\, t \right)}
\end{equation}
being $\beta$ the constant which enters in the definition of the Maxwell's molecules differential cross section, i.e., $\sigma(\|\bv_r\|,\bkhat\cdot \bv_r)=\beta/\|\bv_r\|$.
The distribution function given by Eq.~\eqref{eq:BKW_exact} has physical validity only for $t\geq\bar{t}=(4\pi n_0 \beta)^{-1}6\log{(5/2)}$ 
since it is negative for smaller times.
The three-dimensional velocity space is discretized using $N=\tilde{N}\times \tilde{N}\times\tilde{N}$ discrete velocities. 
Since the BKW distribution function is continuous in the velocity space, the discrete velocities are the quadrature nodes only based on the full-range Hermite polynomial of degree 
$\tilde{N}$. 
A simple first order explicit Euler method is employed to solve Eq.~\eqref{eq:h_galerkin} 
in which transport terms are neglected, $C_{ij\alpha}=0$, 
and the coefficient matrices, Eqs.~\eqref{eq:matrix_B} and~\eqref{eq:matrix_C}, are computed 
using the differential cross section of Maxwell's molecules. 
The initial condition is determined by expanding Eq.~\eqref{eq:BKW_exact} at $t=t_0=\bar{t}$ according to Eq.~\eqref{eq:h_N}.
The ability of the polynomial expansion to reproduce the initial condition is shown in Fig.~\ref{fig:f_initial_condition} where
a comparison is made between the BKW distribution evaluated at $t_0=\bar{t}$ (solid line) and the distribution function as given by Eq.~\eqref{eq:h_N} (dashed lines) 
for different number of expansion polynomials used in each direction of the velocity space $\tilde{N}$. Because of the symmetry of the BKW distribution,
odd full-range Hermite polynomials do not effectively contribute to the expansion defined in Eq.~\eqref{eq:h_N} and thus
only results for the even values of $\tilde{N}$ have been reported. 
A relatively high number of polynomials is required to obtain an accurate representation of the initial condition.
This is reasonable since the BKW solution at $t_0=\bar{t}$ is an highly non-equilibrium distribution function.
However, we notice that even if a low number of expansion polynomials is used, i.e., $N=8$, 
the distribution functions given by Eqs.~\eqref{eq:h_N} and~\eqref{eq:BKW_exact} provide the same mass, momentum and energy.
As a matter of fact, by construction, the first $N$ moments of $h$ and $h_N$ are the same.
In order to quantitatively determine the convergence rate of the polynomial expansion, we report in Fig.~\ref{fig:error_initial_condition}
the error between BGW and the numerical distribution function for two different initial instants of time,
$t_0=\bar{t}$ and $t_0=2\bar{t}$, versus the total number of discrete velocities, $N$.
The error has been computed according to the formula 

\begin{equation}
\label{eq:error}
 \epsilon(t,t_0)=\frac{1}{n_0^2} \int_{\mathbb{R}^3} \left[ f(\bv,t)-f_{N}(\bv,t) \right]^2 \, d\bv
\end{equation}
where $f$ is the BKW distribution function, Eq.~\eqref{eq:BKW_exact}, and $f_N$ is the numerical solution. 
As Figure~\ref{fig:error_initial_condition} clearly shows, for a given $N$, the error at $t_0=2\bar{t}$ is lower than the error at $t_0=\bar{t}$, 
which is not unexpected. Indeed, the higher is $t_0$ the closer is the distribution function to a Maxwellian and therefore the smaller is the number of expansion 
polynomials required to represent the distribution function with a given accuracy. 
Furthermore, independently of the initial instant of time, as $N$ increases, the error decreases with a rate which is higher than any algebraic
order. The proposed velocity space discretization is therefore spectrally accurate. 
In Fig.~\eqref{fig:error_vs_t} the error is reported as a function of time for $\tilde{N}=3, 5$ and $7$ with $t_0=\bar{t}$. 
The most important feature to highlight is that the error asymptotically tends to zero up to machine precision whatever is the value of $\tilde{N}$.
This behavior indicates that distribution function approaches the Maxwellian distribution while conserving mass, momentum and energy.

\subsection{Two dimensional driven cavity flow}
\label{sec:IIIsub2}

\begin{table}[t]
\centering
\begin{tabular}{|l|c|c|c|c|c|c|}
\cline{2-6}
\multicolumn{1}{c|}{}& $\Ma$   & $G/\Ma$         & $D/\Ma$            & $\Theta/\Ma$            & $\Phi_1/\Ma$ \\ \hline 
$\Kn=0.02$ & $\rightarrow 0$ & $0.2340 \pm 0.66\%$ & $0.1741 \pm 2.6\%$ & $0$                    & $0$  \\ \cline{2-6}
            & $0.01$          & $0.2340$          & $0.1741$          & $8.908 \times 10^{-4}$ & $1.142 \times 10^{-3}$ \\ \cline{2-6}
            & $0.1$           & $0.2336$          & $0.1744$          & $8.918 \times 10^{-3}$ & $1.146 \times 10^{-2}$ \\ \cline{2-6}
            & $0.2$           & $0.2323$          & $0.1755$          & $1.778 \times 10^{-2}$ & $2.292 \times 10^{-2}$ \\ \cline{2-6}
            & $0.4$           & $0.2278 \pm 0.47\%$ & $0.1797 \pm 2.13\%$ & $3.518 \times 10^{-2}$ & $4.592 \times 10^{-2}$ \\ \hline \hline
$\Kn=0.1$ & $\rightarrow 0$ & $0.1851 \pm 0.6\%$ & $0.3753 \pm 2.4\%$ & $0$                    & $0$  \\ \cline{2-6}
            & $0.01$          & $0.1851$          & $0.3753$          & $1.745 \times 10^{-3}$ & $1.220 \times 10^{-3}$ \\ \cline{2-6}
            & $0.1$           & $0.1850$          & $0.3756$          & $1.759 \times 10^{-2}$ & $1.324 \times 10^{-2}$ \\ \cline{2-6}
            & $0.2$           & $0.1847$          & $0.3764$          & $3.511 \times 10^{-2}$ & $2.650 \times 10^{-2}$ \\ \cline{2-6}
            & $0.4$           & $0.1839 \pm 0.5\%$ & $0.3796 \pm 1.7\%$ & $6.969 \times 10^{-2}$ & $5.302 \times 10^{-2}$ \\ \hline 
\end{tabular}
\caption{Volume flow rate of the main vortex, $G$, drag coefficient, $D$,
         mean temperature, $\Theta$, and $x_1$-component of mean heat flux, $\Phi_1$,
         along the moving wall, for different values of Knudsen, $\Kn$, and Mach, $\Ma$, numbers.}
\label{tab:results}
\end{table}

\noindent 
The method described in Sec.~\ref{sec:II} is here applied to solve the two-dimensional driven cavity flow of a gas composed 
by hard-sphere molecules. The problem geometry consists of a square enclosure with length $L$.
All the walls are kept at uniform and constant temperature $T_0$.
Initially, the gas is in uniform equilibrium with density $n_0$ and temperature $T_0$. 
The flow is driven by the translation of the lid of the cavity with constant velocity $V_w$~\cite{nv05}. \\
The cavity flow problem has been solved for Mach numbers in the range $[0,0.4]$ and for two values of the Knudsen number, $\Kn=0.02$ and $0.1$.
The Mach and Knudsen numbers are defined, respectively, as $\Ma =V_w/a$ with $a=(5/3 RT_0)^{1/2}$ the speed of sound, and $\Kn=\lambda_0/L$ with
$\lambda_{0}=\mu_{0}/p_{0} (2 RT_{0})^{1/2}$ and $\mu_{0}$ the viscosity of the hard sphere gas~\cite{c88}. 
Numerical results have been obtained by solving Eq.~\eqref{eq:h_galerkin} where the coefficient matrices, Eqs.~\eqref{eq:matrix_A}-\eqref{eq:matrix_C},
are computed with the hard-sphere differential cross section, i.e., $\sigma(\|\bv_r\|,\bkhat\cdot \bv_r)=d^2/4$, being $d$ the molecules' diameter.  
The solution in one time step has been obtained by first integrating the transport equation, Eq.~\eqref{eq:h_galerkin} with $L_{ij}=Q_{ijk}=0$,
and afterwards the space homogeneous equation, Eq.~\eqref{eq:h_galerkin} with $C_{ij\alpha}=0$, using the output of the previous step as initial condition. 
Iterating the process provides the numerical solution at later times. 
The three-dimensional velocity space is discretized using $N=8 \times 8 \times 4$ discrete velocities. 
Because of the presence of solid walls, the discrete velocities along the $x_1$ and $x_2$ directions are the quadrature nodes of the positive and negative
$4th$-order half-range Hermite polynomials whereas discrete velocities along the $x_3$ direction are the the quadrature nodes of the $4th$-order 
full-range Hermite polynomial.
The two-dimensional physical space has been divided into $M=200\times200$ rectangular cells. Their lengths have been uniformly stretched with
a progression ratio $R=0.04$ for $\Kn=0.02$ and $R=0.4$ for $\Kn=0.1$.
The smaller cells are located close to the upper corners of the cavity where severe gradients are anticipated. 
The collision and transport steps have been solved by using simple first-order explicit Euler and upwind schemes.

\begin{figure}[t]
\begin{center}
\includegraphics[width=0.8\textwidth]{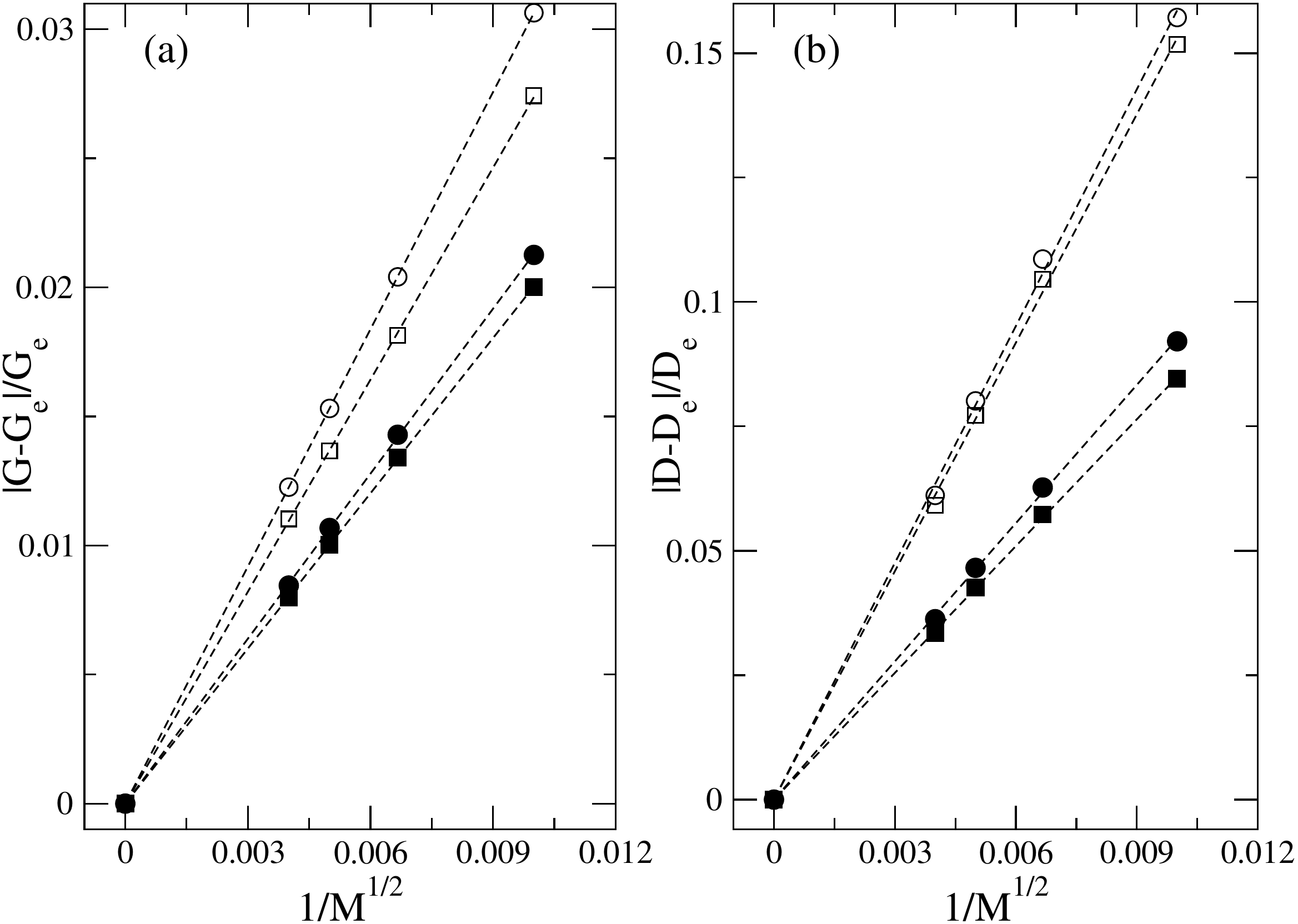}
\caption{Absolute relative error on (a) mean flow rate, $G$, and (b) drag coefficient, $D$,
for two values of the Knudsen number, $\Kn=0.02$ (solid symbols) and $\Kn=0.1$ (empty symbols),
and two values of the Mach number, $\Ma\rightarrow0$ (squares) and $\Ma=0.4$ (circles),
versus the size of the grid in the physical space, $1/M^{1/2}$.
Dashed lines are the least-mean square fit of the results.}
\label{fig:convergence}
\end{center}
\end{figure}

\begin{figure}[t]
 \centering
 {\includegraphics[width=0.45\textwidth]{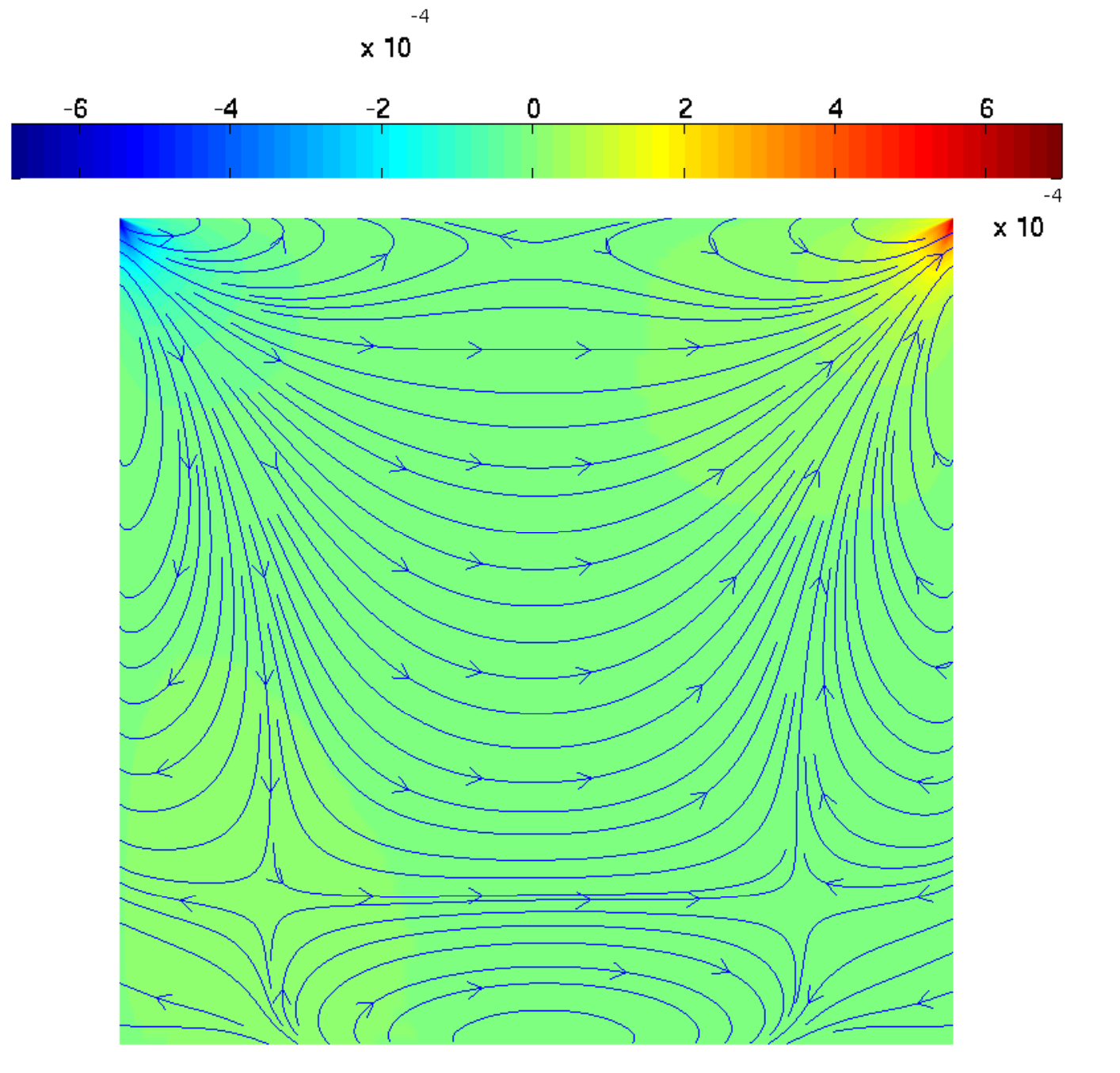}} \quad
 {\includegraphics[width=0.45\textwidth]{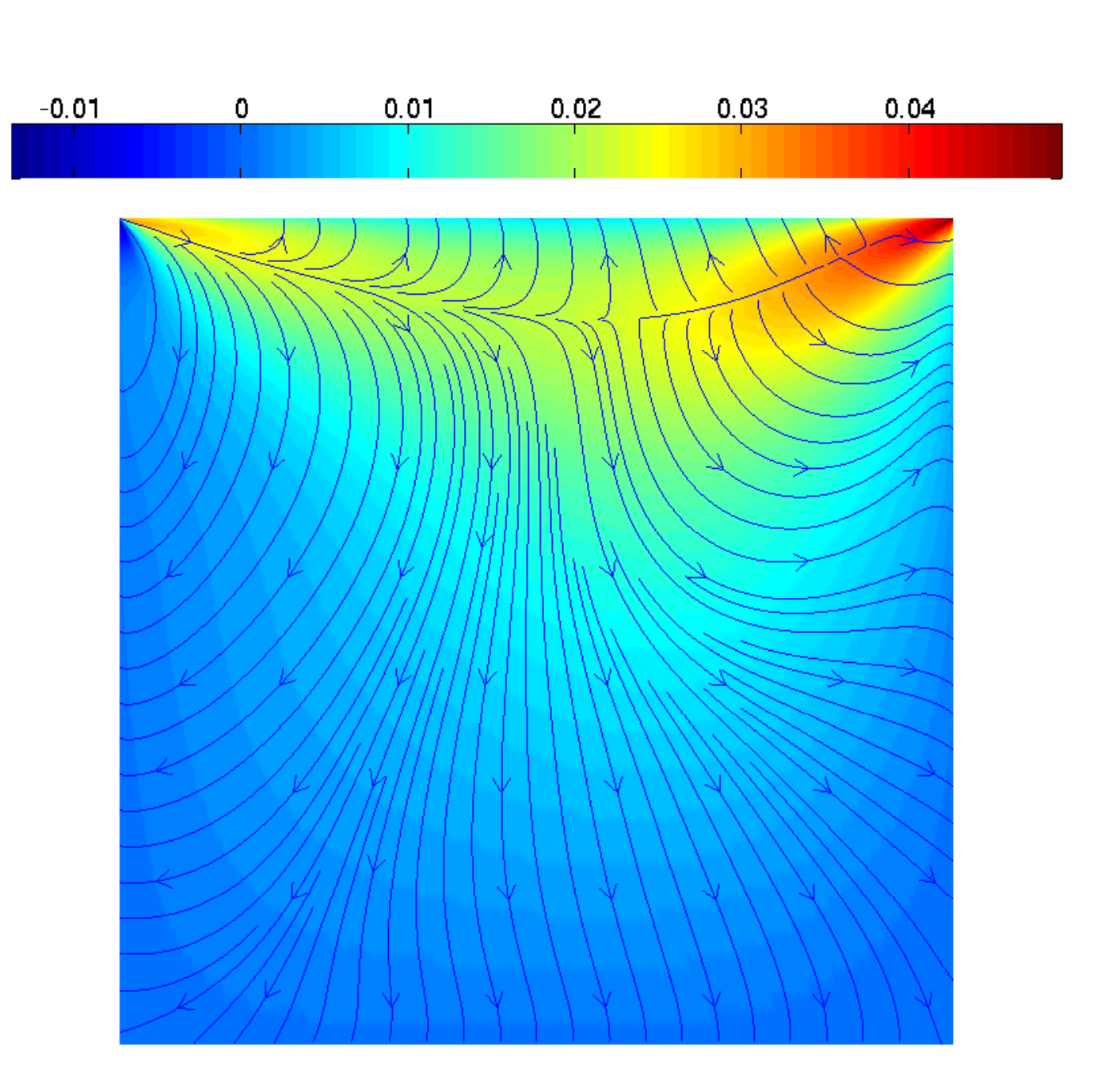} }
 \caption{Snapshots of heat flux lines overlaid on the temperature field $(T-T_0)/T_0$ at $\Kn=0.02$ and
          $\Ma=0.01$ (left panel) and $\Ma=0.4$ (right panel).$\Kn=0.02$}
 \label{fig:fields}
\vspace{1.0cm}
\end{figure}

\begin{figure}[t]
\begin{center}
\includegraphics[width=0.8\textwidth]{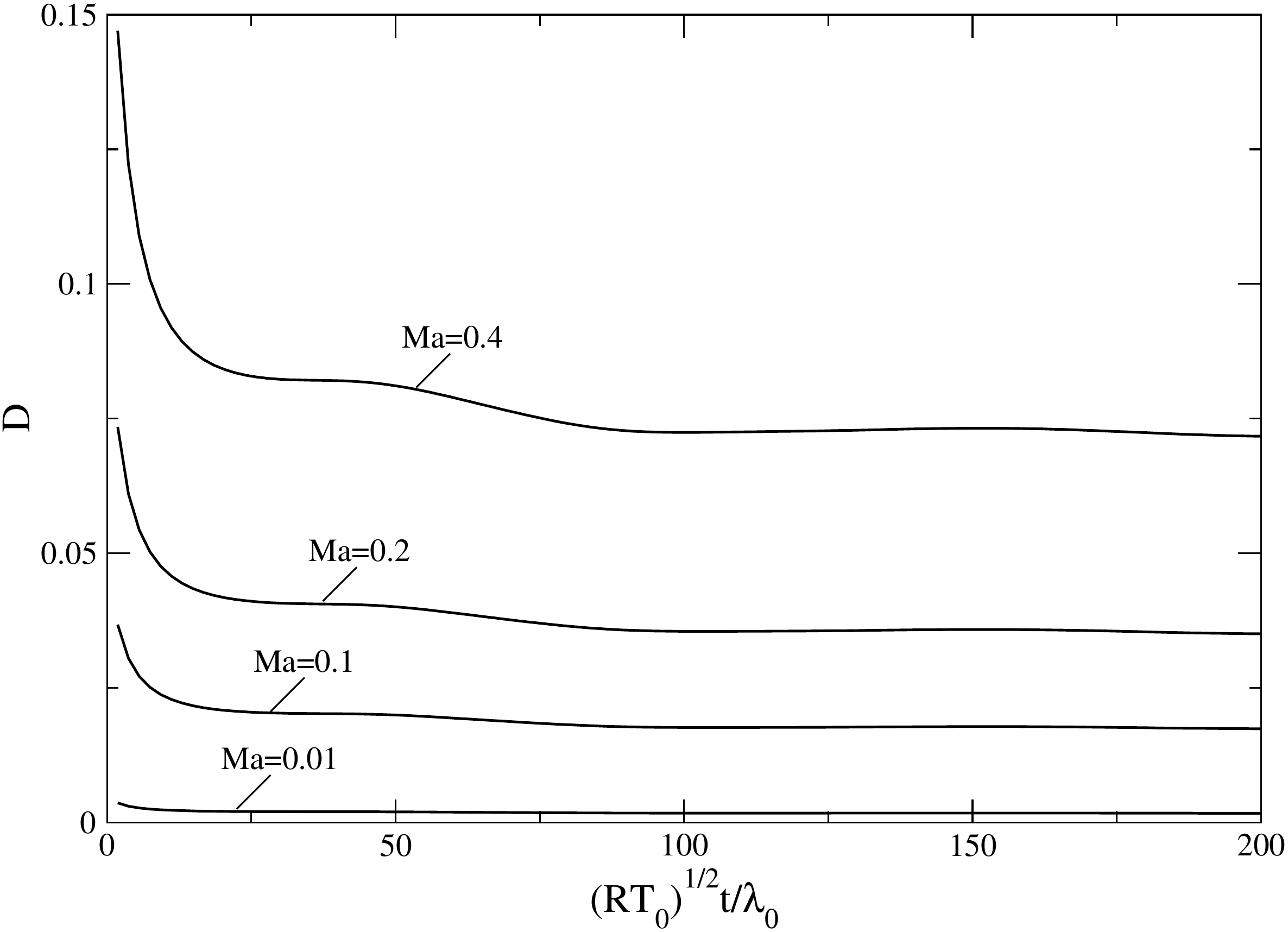}
\caption{Drag coefficient versus dimensionless time for different Mach numbers, $\Ma$. $\Kn=0.02$.}
\label{fig:drag}
\end{center}
\end{figure}

\noindent
We firstly establish the convergence rate of the method by computing two global flow field
properties, namely the drag coefficient, $D$,
and the flow rate of the main vortex, $G$. We refer to Refs.~\cite{fgg11b,ggdi12} for precise 
definitions of all the macroscopic quantities used in the present Section. 
The absolute relative error in the long-term dimensionless values of $D$ and $G$ are
shown in Figs~\ref{fig:convergence}a and \ref{fig:convergence}b, respectively,
versus the dimensionless spatial grid size, $1/\sqrt{M}$,
for two values of the Knudsen number, $\Kn=0.02$ (solid symbols) and $\Kn=0.1$ (empty symbols),
and two values of the Mach number, $\Ma\rightarrow0$ (squares) and $\Ma=0.4$ (circles).
It is worth noticing that the scheme which has been used for numerically solving the Boltzmann equation in the linearized regime, that is for $\Ma\rightarrow0$,
is provided by Eq.~\eqref{eq:h_galerkin} in which the second term on the right hand side is disregarded.
The exact values of $D$ and $G$, which are referred to as $D_{e}(\Ma)$ and
$G_{e}(\Ma)$, have been extrapolated from the linear fit of the results
when $1/\sqrt{M} \rightarrow 0$.
The linear behavior of the absolute relative errors demonstrates that
the results are in the asymptotic range of convergence and the method is overall first
order accurate in the physical space~\cite{s06}. Table~\ref{tab:results} shows the
numerical predictions of the volume flow rate of the main vortex $G$, the drag coefficient, $D$,
the mean temperature, $\Theta$, and the $x_1$-component of the mean heat flux, $\Phi_1$, along the moving wall.
The error bound for $G$ and $D$ is also reported for $\Ma\rightarrow0$ and $\Ma=0.4$.
The largest percentage errors in $D$ and $G$ are about $2.6\%$ and $0.66\%$ and they 
are attained at $\Kn=0.02$ in the linearized regime. Results for intermediate Mach numbers are expected to retain a similar level of accuracy.
\begin{figure}[t]
\centering
\includegraphics[width=0.8\textwidth]{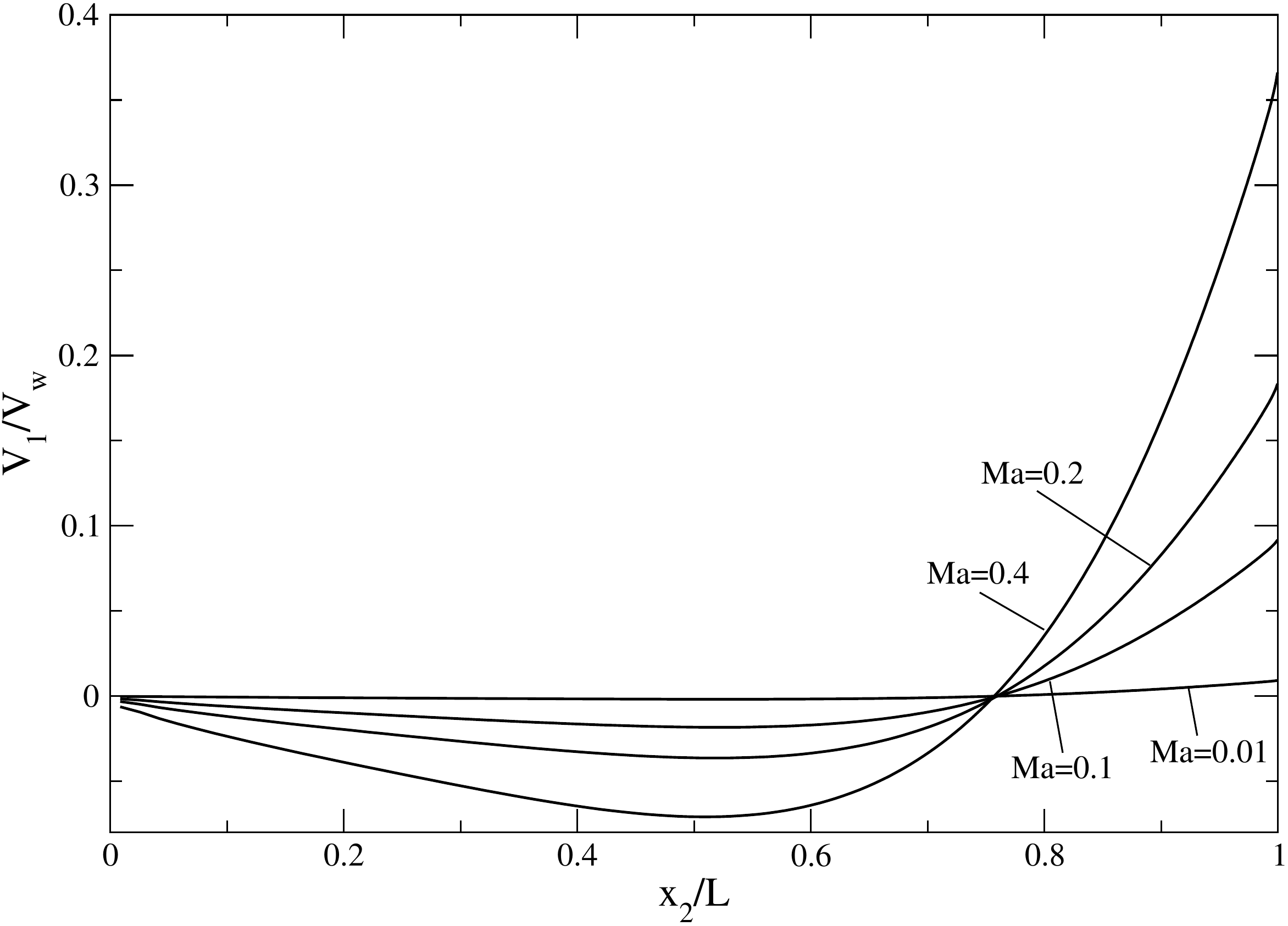} 
\caption{Profiles of the dimensionless horizontal mean velocity along the vertical
         line crossing the center of the cavity, $V_{1}/V_{w}$ for $\Kn=0.02$ and different Mach numbers.}
\label{fig:profili_a}
\vspace{1.2cm}
\end {figure}
\begin{figure}[t]
\centering
\includegraphics[width=0.8\textwidth]{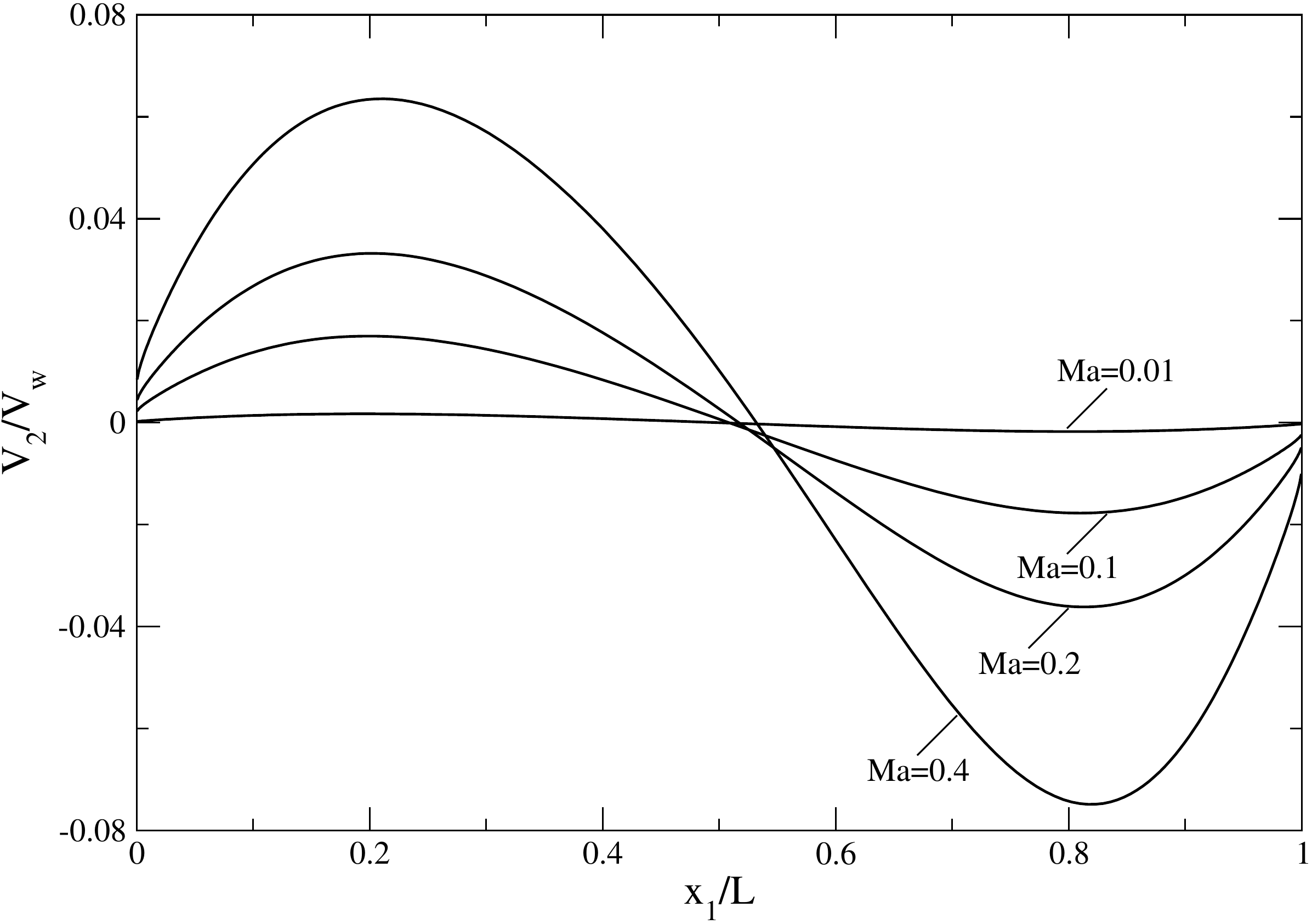}
\caption{Profiles of the dimensionless vertical mean velocity
         component along the horizontal line crossing the center of the cavity for $\Kn=0.02$ and different Mach numbers.}
\label{fig:profili_b}
\end {figure}
Figures \ref{fig:profili_a} and \ref{fig:profili_b} show the profiles of the
dimensionless horizontal component of the velocity, $V_{1}/V_{w}$, along the vertical line
crossing the center of the cavity and the dimensionless vertical component of the velocity, $V_{2}/V_{w}$,
along the horizontal line crossing the center of the cavity, respectively, 
for $\Kn=0.02$ and Mach numbers in the range $[0,0.4]$.
The possibilities of the proposed method of solution are further illustrated by the computations reported in Figs.~\ref{fig:fields} and~\ref{fig:drag}. 
Fig.~\ref{fig:fields} shows snapshots of the heat flux lines superimposed to the temperature field $(T-T_0)/T_0$ for two values of the Mach number, $\Ma=0.01$ (left panel) and
$\Ma=0.4$ (right panel) whereas in Fig.~\ref{fig:drag} the time evolution of the drag coefficient, $D$, is reported during the simulation for $\Kn=0.02$ 
and Mach numbers in the range $[0,0.4]$.
These results would be difficult to obtain with a particle-based method since computationally expensive time and/or ensemble averaging are needed to provide such smooth 
macroscopic fields.

\noindent
We conclude with a few remarks on the parallel implementation of the proposed method of solution.  
Profiling of the serial code has highlighted that the collision step has a computational cost which is four order of magnitude greater 
than the computational cost of the streaming step and it takes about $99\%$ of the total simulation time.
Therefore, only the collision step has been parallelized leaving the transport step performed by all MPI processes.
Matrix-vector operations involved in the collision step are independent along $i$, $j$ and $k$ indices. 
As parallelization strategy we decided to implement an hybrid solution decomposing the simulation along the $i$ index with OpenMP and the $k$ index with MPI. 
Simulations have been performed on a cluster which is comprised by $128$ nodes, each node having $12$ cores Intel Xeon X5660 and $24\,\mbox{GB}$ of RAM. 
The nodes are interconnected by an Infiniband QDR/DDR Voltaire network. 
Benchmarks showed that the parallelization strategy which has been adopted yields an efficiency greater than $70\%$ up to 128 cores.
The results reported in this Section were obtained using 64 cores, with a memory requirement for each simulation of about $100\,\mbox{MB}$ of RAM and a computational 
cost of the order of $O(M\times N^3)$ floating point operations per time iteration, being $M=400000$ and $N=256$ the total
numbers of cells in the physical and velocity space, respectively. The time required to execute one time iteration was about $13\,s$.

\section{Conclusion}
\label{sec:IV}

We developed a deterministic method of solution for the Boltzmann equation based on a pseudo-spectral Galerkin discretization of the velocity space.
The novel aspect of the proposed method is the choice of trial and test functions, namely
the collocation basis functions related to weights and roots of a Gauss-Hermite quadrature formula defined on the basis of half- and/or full-range Hermite polynomials, 
Eqs.~\eqref{eq:appendix_basis}. 
The use of half-range Hermite polynomials is an important feature of the method since it permits a straightforward and exact formulation of boundary
conditions.
Due to this discretization of the velocity space, the Boltzmann equation simplifies to a system of hyperbolic conservation laws with 
non-linear source terms in the physical space, Eqs.~\eqref{eq:h_galerkin}. A fully discrete numerical scheme is derived by a first-order time splitting
of the evolution operator which couples a first-order finite-volume scheme for the transport step with a first-order explicit Euler scheme for the relaxation step. 
However, more sophisticated methods can be used for coupling and solving both transport and relaxation steps.
Numerical results have been presented for the space homogeneous relaxation to equilibrium of Maxwell molecules
and the two-dimensional cavity flow of a gas composed by hard-sphere molecules.
Overall, the numerical results are affected by an error of less than a few percent. \\ 
The proposed method of solution has a number of attracting features such as explicit identification and exact treatment of boundary conditions,
spectral accuracy in the velocity space and preservation of the collision invariants.
While the method can in principle handle a wide range of Knudsen and Mach numbers, it turns out to be feasible 
only in dealing with slightly rarefied low Mach number gas flows. 
The main obstacle is the large number of operations required to evaluate the collision integral.
However, a preliminary study indicates that the numerical code possesses very good scalability 
on a memory distributed cluster of GPUs and supercomputing architectures like IBM Blue Gene/Q.
A throughly discussion of these parallel implementations will be the subject of a future publication.

\section{Acknowledgments}
The authors would like to thank Prof. A. Frezzotti for several helpful discussions and for critically reading the paper.
Moreover, they gratefully acknowledge the support received from Regione Lombardia and SAES Getters within the framework of the
projects ``Modelli Matematici per la Progettazione di Barriere Attive (MOBA)''.
This work has also been supported by  Regione Lombardia and CILEA Consortium through a LISA Initiative (Laboratory for Interdisciplinary Advanced
Simulation) 2011 grant [link:http://lisa.cilea.it].

\section*{Appendix: Half-range Gauss-Hermite quadrature}
\begin{figure}[t]
 \centering 
 \includegraphics[width=0.8\textwidth]{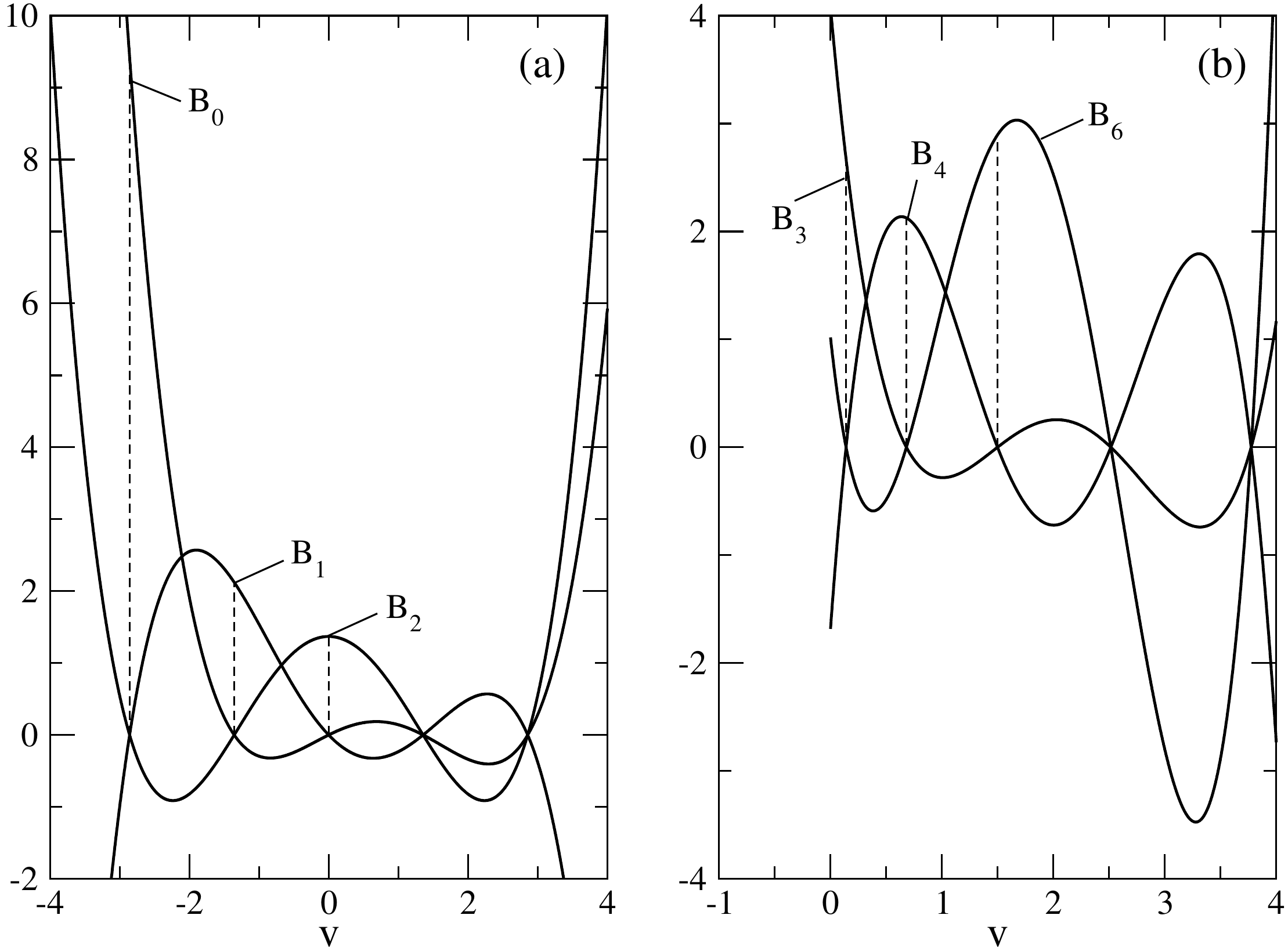}
 \caption{Collocation basis function based on (a) full-range Hermite polynomials, $N=3$ and 
                                              (b) positive half-range Hermite polynomials, $N=6$.}
 \label{fig:basi_cd}
\end{figure}

We introduce
polynomials in three variables defined as the Cartesian product of polynomials in one variable and we organize them
according to the lexicographical order, i.e., 
$\tilde{\mathcal{P}}_i(\bv)=\tilde{\mathcal{P}}^{(1)}_{i_{1}}(v_{1})\tilde{\mathcal{P}}^{(2)}_{i_{2}}(v_{2})\tilde{\mathcal{P}}^{(3)}_{i_{3}}(v_{3})$
where $i=i_1+\tilde{N}_1 i_2+\tilde{N}_1 \tilde{N}_2 i_3$, 
$i_\alpha=0,\ldots,\tilde{N}_\alpha-1$ with $\alpha=1,2,3$ and $\tilde{N}_\alpha$
the number of polynomials in the $v_\alpha$ velocity component.  
By applying the Gram-Schmidt procedure to the dimensionless monomial powers $1,v_\alpha/(RT_0)^{1/2},v_\alpha^2/(RT_0),\ldots$,
polynomials $\tilde{\mathcal{P}}^{(\alpha)}_{i_\alpha}$ can be constructed that are orthonormal in the range $[a_{\alpha},b_{\alpha}]$ with respect 
to the weighted inner product defined as

\begin{equation}
\label{eq:ap_inner}
\int_{a_\alpha}^{b_\alpha} 
M(v_{\alpha}) \tilde{\mathcal{P}}^{(\alpha)}_{i_\alpha}(v_\alpha) \tilde{\mathcal{P}}^{(\alpha)}_{j_\alpha}(v_\alpha) \; dv_\alpha = \delta_{i_\alpha,j_\alpha}
\end{equation}
where $\delta_{i_\alpha,j_\alpha}$ denotes the Kronecker delta function which is $1$ for $i_\alpha=j_\alpha$ and $0$ otherwise.
In Eq.~\eqref{eq:ap_inner}, $M(v_\alpha)$ is the Maxwellian weight function with unit mass which reads

\begin{equation}
 M(v_\alpha) = \frac{1}{\sqrt{\pi RT_0}} \exp{\left( -\frac{v^2_\alpha}{2RT_0} \right)}
\end{equation}
As a matter of fact, polynomials $\tilde{\mathcal{P}}^{(\alpha)}_{i_\alpha}$ can be regarded as defined over the whole real axis by setting their values to zero
outside the interval $[a_{\alpha},b_{\alpha}]$. \\
The polynomials afore introduced permit to define the following quadrature formula

\begin{equation}
\label{eq:appendix_quadrature}
\int_{\mathbb{R}^{3}} M(\bv) f(\bv) \; d\bv   
\simeq \sum_{i_1=0}^{\tilde{N}_1-1} \sum_{i_2=0}^{\tilde{N}_2-1} \sum_{i_3=0}^{\tilde{N}_3-1} \tilde{w}^{(1)}_{i_1} \tilde{w}^{(2)}_{i_2} \tilde{w}^{(3)}_{i_3} 
                                                                      f(\tilde{z}^{(1)}_{i_1},\tilde{z}^{(2)}_{i_2},\tilde{z}^{(3)}_{i_3}) 
\end{equation}
where $\tilde{z}^{(\alpha)}_{i_\alpha}$ are the roots of the polynomial $\tilde{\mathcal{P}}^{(\alpha)}_{N_\alpha}$ and 
$\tilde{w}^{(\alpha)}_{i_\alpha}$ are quadrature weights which are determined by solving the linear systems of equations

\begin{equation}
\label{eq:appendix_weights}
 \sum_{j_\alpha=0}^{\tilde{N}_\alpha-1} \tilde{\mathcal{P}}^{(\alpha)}_{i_\alpha}(\tilde{z}^{(\alpha)}_{j_{\alpha}}) \tilde{w}^{(\alpha)}_{j_{\alpha}} = \delta_{0,i_\alpha}, 
\; \; \; i_\alpha=0,\ldots,N_\alpha-1
\end{equation}
where $\alpha=1,2,3$.
The quadrature formula given by Eq.~\eqref{eq:appendix_quadrature} can be shown to provide the correct result 
for the integral of functions that are products of polynomials in each velocity component $v_\alpha$ up to degree $2\tilde{N}_\alpha-1$,
as long as their support is contained in $[a_1,b_1]\times[a_2,b_2]\times[a_3,b_3]$. \\
In the present work, three polynomial sets are used that satisfy Eq.~\eqref{eq:ap_inner} in a different interval $[a_\alpha,b_\alpha]$. 
The support of full-range Hermite polynomials, $H^f_{i_\alpha}$, is $(a_\alpha,b_\alpha)=(-\infty,+\infty)$
whereas the support of negative, $H^{n}_{i_\alpha}$, and positive, $H^{p}_{i_\alpha}$, half-range Hermite polynomials are $(a_\alpha,b_\alpha)=(-\infty,0)$ and 
$(a_\alpha,b_\alpha)=(0,+\infty)$, respectively.
Zeros and weigths of $H^f_{i_\alpha}, H^{n}_{i_\alpha}$ and $H^{p}_{i_\alpha}$ are referred to as $z^{f}_{i_\alpha}, z^{n}_{i_\alpha}, z^{p}_{i_\alpha}$ 
and $w^{f}_{i_\alpha}, w^{n}_{i_\alpha}, w^{p}_{i_\alpha}$, respectively.
It is not difficult to prove that the zeros of full-range Hermite polynomials are symmetric
about the zero velocity, i.e., $z^f_{i_\alpha}=z^f_{\tilde{N}_\alpha-1-\alpha}$, 
whereas the zeros of positive and negative half-range Hermite polynomials have equal magnitude but different sign, i.e., $z^m_{i_\alpha}=-z^p_{i_\alpha}$.\\
The collocation basis functions introduced in Section~\ref{sec:II} are
based on roots and weights of the Gauss-Hermite quadrature formula defined by Eq.~\eqref{eq:appendix_quadrature} for the polynomial set 
$\mathcal{P}^{(\alpha)}_{i_\alpha}$ . 
More precisely, the collocation basis functions read

\begin{multline}
\label{eq:appendix_basis}
\mathcal{B}_i (v_1,v_2,v_3) = \sum_{j_{1}=0}^{N_1-1} \sum_{j_{2}=0}^{N_2-1} \sum_{j_{3}=0}^{N_3-1} 
                              \sqrt{w^{(1)}_{i_1} w^{(2)}_{i_2} w^{(3)}_{i_3} }  
                              \mathcal{P}^{(1)}_{j_{1}} (z^{(1)}_{i_1}) \mathcal{P}^{(2)}_{j_{2}} (z^{(2)}_{i_2}) \mathcal{P}^{(3)}_{j_{3}} (z^{(3)}_{i_3}) \\
                              \mathcal{P}^{(1)}_{j_{1}} (v_1) \mathcal{P}^{(2)}_{j_{2}} (v_2) \mathcal{P}^{(3)}_{j_{3}} (v_3)
                              \; \; \; 
\end{multline}
where $z^{(\alpha)}_{i_\alpha}$ and $w^{(\alpha)}_{i_\alpha}$
are the nodes and weights of the quadrature formula given by Eq.~\eqref{eq:appendix_quadrature} based on the polynomials
$\mathcal{P}^{(\alpha)}_{N_\alpha}$.
These are defined differently depending whether or not the deviational part of the distribution function is expected to have 
a discontinuity in the $v_\alpha$ velocity component. More precisely, if $h$ is continuous in the velocity space
then $\mathcal{P}^{(\alpha)}_{i_\alpha} = H^f_{i_\alpha}$ and $z^{(\alpha)}_{i_\alpha}=z^f_{i_\alpha}$, $w^{(\alpha)}_{i_\alpha}=w^f_{i_\alpha}$.
Instead, if $h$ presents a discontinuity in the $v_\alpha$ velocity component then 
\begin{equation}
 \mathcal{P}^{(\alpha)}_{i_\alpha} =
 \begin{cases}
   H^{n}_{i_\alpha} & i_\alpha=0,\ldots,N_\alpha/2-1 \\
   H^{p}_{i_\alpha-N_\alpha/2} & i_\alpha= N_\alpha/2,\ldots,N_\alpha-1
 \end{cases}
\end{equation}
Likewise
\begin{equation}
 z^{(\alpha)}_{i_\alpha} =
 \begin{cases}
   z^{n}_{i_\alpha} & i_\alpha=0,\ldots,N_\alpha/2-1 \\
   z^{p}_{i_\alpha-N_\alpha/2} & i_\alpha= N_\alpha/2,\ldots,N_\alpha-1
 \end{cases}
\end{equation}
and
\begin{equation}
 w^{\alpha}_{i_\alpha} =
 \begin{cases}
   w^{m}_{i_\alpha} & i_\alpha=0,\ldots,N_\alpha/2-1 \\
   w^{p}_{i_\alpha-N_\alpha/2} & i_\alpha= N_\alpha/2,\ldots,N_\alpha-1
 \end{cases}
\end{equation}
Equation~\eqref{eq:appendix_basis} can be rewritten with a more compact notation as follows
\begin{equation}
\label{eq:appendix_basis_compact}
\mathcal{B}_i (\bv) = \sum_{j=0}^{N-1} \sqrt{w_i} \mathcal{P}_j (\bz_i) \mathcal{P}_j (\bv) 
\end{equation}
By using the quadrature formula given by Eq.~\eqref{eq:appendix_quadrature}, it is possible to prove that the collocation basis functions defined by Eq.~\eqref{eq:appendix_basis}
are orthonormal with respect to the Maxwellian weight factor,

\begin{equation}
 \int_{\mathbb{R}^3} M(\bv) \mathcal{B}_i (\bv) \mathcal{B}_j (\bv) d\bv= \delta_{i,j}
\end{equation}
and that the $i$-th basis function vanishes on all nodes but node $i$

\begin{equation}
 \label{eq:appendix_collocation_property}
 \mathcal{B}_i (\bz_j)=\delta_{ij}/\sqrt{w_i}
\end{equation}
For illustrative purposes, 
in Figs~\ref{fig:basi_cd}, we report the collocation basis functions in one-variable based on the quadrature nodes of the full-range Hermite polynomials (left panel)
and positive half-range Hermite polynomials (right panel).  

% ==========================================================================================================


\begin{thebibliography}{50}
\bibitem{c88} C. Cercignani,
              The Boltzmann Equation and Its Applications,
                Springer-Verlag, New Yor1Gk, 1988.
\bibitem{b94} G. A. Bird,
              Molecular Gas Dynamics and the Direct Simulation of Gas
              Flows,
              Oxford University Press, 1994.
\bibitem{rw96} S. Rjasanow, W. Wagner,
               ``A stochastic weighted particle method for the Boltzmann equation'',
               J. Comput. Phys. 124 (1996) 243-253.
\bibitem{hh07} T. M. M. Homolle , N. G. Hadjiconstantinou,
               ``A low-variance deviational simulation Monte Carlo for the Boltzmann equation'',
               J. Comput. Phys. 226 (2007) 2341-2358.
\bibitem{ar01} V. V. Aristov, 
               {\em Direct Methods for Solving the Boltzmann Equation
                    and Study of Nonequilibrium Flows}, Kluwer Academic Publishers, 2001.
\bibitem{dpr04} P. Degond, L. Pareschi and G. Russo,
                {\em Modeling and Computational Methods for Kinetic Equations},
                Series: Modeling and Simulation in Science, Engineering and Technology,
                Birkh\"auser, 2004. 
\bibitem{lgffc07} S. Lorenzani, L. Gibelli, A. Frezzotti, A. Frangi, C. Cercignani,
                  ``Kinetic approach to gas flows in microchannels'',
                  Nanos. Micros. Therm. 11 (2007) 211-226.
\bibitem{cffggl08} C. Cercignani, A. Frangi, A. Frezzotti, G. P. Ghiroldi, L. Gibelli, and S. Lorenzani, 
                 ``On the application of the Boltzmann equation to the simulation of fluid structure interaction in micro-electro-mechanical-systems'',
                  Sens. Lett. 6 (1) (2008) 121-129.
\bibitem{ar11} V. V. Aristov, O. I. Rovenskaya,
               ``Application of the Boltzmann kinetic equation to the eddy problems'',
                Comput. Fluids. 50 (2011) 189-198.
\bibitem{sd04} Y. Sone, T. Doi,
               ``Ghost effect of infinitesimal curvature of in the plane Couette flow of a gas in the continuum limit'', 
               Phys. Fluids 16 (2004) 952-971.
\bibitem{s02} Y. Sone,
              {\em Kinetic Theory and Fluid Dynamics}
              (Birkh\"auser, Boston, 2002). 
\bibitem{ddm10} P. Degond, G. Dimarco, L. Mieussens,
                ``A multiscale kinetic-fluid solver with dynamic localization of kinetic effects'',
                J. Comput. Phys. 229 (2010) 4907-4933.
\bibitem{ap11} A. Alaia, G. Puppo,
               ``A hybrid method for hydrodynamic-kinetic flow. Part I: A particle-grid method for reducing stochastic noise in kinetic regimes'',
                J. Comput. Phys. 230 (2011) 5660-5683. 
\bibitem{fgg11a} A. Frezzotti, G. P. Ghiroldi, L. Gibelli,
                 ``Solving model kinetic equations on GPUs'',
                  Comput. Fluids. 50 (2011) 136-146.
\bibitem{fgg11b} A. Frezzotti, G. P. Ghiroldi, L. Gibelli,
                 ``Solving the Boltzmann equation on GPUs'',
                 Comput. Phys. Commun. 182 (2011) 2445-2453.
\bibitem{hg12} J. R. Haack, I. M. Gamba,
               ``High performance computing with a conservative spectral Boltzmann solver'',
                28th International Symposium on Rarefied Gas Dynamics, 9-13 July 2012, Zaragoza, Spain, AIP Conference
               Proceedings American Institute of Physics, 1501, 334-341 (2012).
\bibitem{m00} L. Mieussens,
              ``Discrete-velocity models and numerical schemes for the Boltzmann-BGK equation in plane and axisymmetric geometries'',
              J. Comput. Phys. 162 (2000) 429-466. 
\bibitem{a11} A. M. Alekseenko,
              ``Numerical properties of high order velocity solutions to the BGK kinetic equation'',
              App. Num. Math. 61 (2011) 410-427.
\bibitem{l02} R. J. LeVeque,
              {\em  Finite Volume Methods for Hyperbolic Problems},
              Cambridge University Press, 2002.
\bibitem{ph02} V. A. Panferov, A. G. Heintz, 
               ``A new consistent discrete-velocity model for the Boltzmann equation'',
               Math. Method. Appl. Sci. 25 (7) (2002) 571-593.
\bibitem{mpr12} C. Mouhot, L. Pareschi, T. Rey,
                ``Convolutive decomposition and fast summation methods for discrete-velocity approximations
                  of the Boltzmann equation'', arxiv:1021.3986v2 
\bibitem{pr00} L. Pareschi, G. Russo,
               ``Numerical solution of the Boltzmann equation. I. Spectrally accurate approximation of
                 the collision operator'',
               SIAM J. Numer. Anal. 37 (4) (2000) 1217-1245.
\bibitem{mp04} C. Mouhot, L. Pareschi,
             ``Fast algorithms for computing the Boltzmann collision operator'',
             C. R. Acad. Sci. Paris S\'er I Math 339 (1) (2004) 71-76.             
\bibitem{fmp06} F. Filbet, C. Mouhut and L. Pareschi,
                ``Solving the Boltzmann equation in $Nlog_2N$''
               SIAM J. Sci. Comput. 28 (3) (2006) 1029-1053.
\bibitem{gt09} I. M. Gamba, S. H. Tharkabhushanam,
               ``Spectral-Lagrangian methods for collisional models of non-equilibrium statistical states'',
               J. Comput. Phys. 228 (2009) 2012-2036.
\bibitem{bh08} L. L. Baker, N. G. Hadjiconstantinou,
               ``Variance-reduced Monte Carlo solutions of the Boltzmann equation for low-speed gas flows: A discontinuous
                 Galerkin formulation'',
               Int. J. Numer. Meth. Fl. 58 (2008) 381-402.
\bibitem{m11} A. Majorana,
              ``A numerical model of the Boltzmann equation related to the discontinuous Galerkin method'',
               Kin. Rel. Mod. 4 (1) (2011) 139-151. 
\bibitem{aj12} A. Alekssenko, E. Josyula, 
               ``Deterministic solution of the Boltzmann equation using a discontinuous Galerkin velocity discretization'',
               28th International Symposium on Rarefied Gas Dynamics, 9-13 July 2012, Zaragoza, Spain, AIP Conference
               Proceedings American Institute of Physics, 1501, pp. 279-286 (2012).
\bibitem{aj12_b} A. Alekssenko, E. Josyula, 
               ``Deterministic solution of the Boltzmann equation using discontinuous Galerkin discretization in velocity space'',
               airxiv:1301.1099v1
\bibitem{o92} T. Ohwada,
              ``Structure of normal shock wave: Direct numerical analysis of the Boltzmann equation for hard-sphere molecules'',
              Phys. Fluids 5 (1) (1992) 217-234.
\bibitem{soa89} Y. Sone, T. Ohwada, K. Aoki, 
              ``Temperature jump and Knudsen layer in a rarefied gas over a plane wall: Numerical analysis of the linearized
                Boltzmann equation for hard-sphere molecules'', 
              Phys. Fluids 1 (1989) 363-370.
\bibitem{gwc06} M. K. Gobbert, S. G. Webster and T. S. Cale,
                ``A Galerkin method for the simulation of the transient 2-D/2-D and 3-D/3-D
                  linear Boltzmann equation'',
                J. Sci. Comput. 30 (2) (2007) 237-273.
\bibitem{gjz57} E. P. Gross, E. A. Jackson, S. Ziering,
                ``Boundary value problem in kinetic theory of gases'',
                Ann. Phys. 1 (1957) 141-167.
\bibitem{s03} C. E. Siewert,
             ``The linearized Boltzmann equation: Concise and accurate
               solutions to basic flow problems'',
              Z. Angew. Math. Phys. 54 (2003) 273-303.
\bibitem{fgf09} A. Frezzotti, L. Gibelli, B. Franzelli,
                ``A moment method for low speed flows'',
                Cont. Mech. Thermodyn. 21 (6) (2009) 495-509.
\bibitem{g12} L. Gibelli,
              ``Velocity slip coefficients based on the hard-sphere Boltzmann equation'',
              Phys. Fluids {\bf 24} (2012) 022001.
\bibitem{ls83} M. J. Lindenfeld, B. D. Shizgal,
               ``The Milne problem: A study of the mass dependence'',
               Phys. Rev. A 27 (3) (1983) 1657-1670.
\bibitem{lz04} Z. Li, H. Zhang,
               ``Gas kinetic algorithm using Boltzmann model equation'',
               Computers \& Fluids 33 (2004) 967–991.
\bibitem{fs01} J. Fan and C. Shen,
               ``Statistical simulation of low-speed rarefied gas flows'',
               J. Comput. Phys. {\bf 167} (2001) 393-412.
\bibitem{b75} A. V. Bobylev,
              ``Exact solutions of the Boltzmann equation'',
              Dokl. Akad. Nauk. S. S. S. R. 225 (1975) 1296-1299 (Russian).
\bibitem{b88} A. V. Bobylev, 
              ``The theory of the nonlinear spatially uniform Boltzmann equation for Maxwell molecules'',
              Math. Phys. Reviews 7 (1988) 111–233.
\bibitem{kw77} M. Krook, T. T. Wu,
               ``Exact solutions of the Boltzmann equation'',
               Phys. Fluids 20 (10) (1977) 1589-1595.
\bibitem{fr03} F. Filbet, G. Russo,
               ``High order numerical methods for the space non homogeneous Boltzmann equation'',
               J. Comput. Phys. 186 (2003) 457-480.
\bibitem{nv05} S. Naris, D. Valougeorgis,
               ``The driven cavity flow over the whole range of the Knudsen number'',
               Phys. Fluids 17 (2005) 097106.
\bibitem{mesbr07} S. Mizzi, D. R. Emerson, S. K. Stefanov, R. W. Barber, J. M. Reese, 
                  ``Effects of Rarefaction on Cavity Flow in the Slip Regime”, 
                 J. Comput. Theor. Nano. 4 (4) (2007) 817-822.
\bibitem{ggdi12} G. P. Ghiroldi, L. Gibelli, P. Dagna, A. Invernizzi,
               ``Linearized Boltzmann Equation: A Preliminary Exploration of its Range of Applicability'',
               28th International Symposium on Rarefied Gas Dynamics, 9-13 July 2012, Zaragoza, Spain, AIP Conference
               Proceedings American Institute of Physics, 1501, pp. 735-741 (2012).
\bibitem{s06} M. D. Salas,
              ``Some observations on grid convergence'',
               Comp. Fluids 35 (2006) 688-692.
\end{thebibliography}
\end{document}